\newif\iffigs
\figstrue
\iffigs
\fi
\def\drawing #1 #2 #3 {
\begin{center}
\setlength{\unitlength}{1mm}
\begin{picture}(#1,#2)(0,0)
\put(0,0){\framebox(#1,#2){#3}}
\end{picture}
\end{center} }

\documentclass[14pt,aps,pre,twocolumn,floatfix,twoside,superscriptaddress,showpacs,amsfonts,amssymb,amsmath,preprintnumbers]{revtex4}

\usepackage{amsmath}
\usepackage{epsfig}
\usepackage{epsf}
\usepackage{graphicx}

\begin{document}
\title{Dynamo Onset as a First-Order Transition: Lessons from a Shell Model for Magnetohydrodynamics}
\author{Ganapati Sahoo}
\email{ganapati@physics.iisc.ernet.in}
\affiliation{Centre for Condensed Matter Theory,
Department of Physics, Indian Institute of Science, Bangalore 560012, India.}
\author{Dhrubaditya Mitra}
\email{dhruba.mitra@gmail.com}
\affiliation{Astronomy Unit, School of Mathematical Sciences, Queen Mary College, University of London, London, E1 4NS, UK}
\author{Rahul Pandit}
\email{rahul@physics.iisc.ernet.in}
\altaffiliation{\\ Also at Jawaharlal Nehru Centre for Advanced 
Scientific Research, Bangalore, India.}
\affiliation{Centre for Condensed Matter Theory, Department of Physics,
 Indian Institute of Science, Bangalore 560012, India.}
\begin{abstract}
We carry out systematic and high-resolution studies of dynamo action in 
a shell model for magnetohydrodynamic (MHD) turbulence over wide ranges 
of the magnetic Prandtl number $Pr_{\rm M}$ and the magnetic Reynolds
number $Re_{\rm M}$. Our study suggests that it is natural to think 
of dynamo onset as a nonequilibrium, first-order phase transition
between two different turbulent, but statistically steady, states.
The ratio of the magnetic and kinetic energies is a convenient
order parameter for this transition. By using this order parameter, we 
obtain the stability diagram (or nonequilibrium phase diagram) 
for dynamo formation in our MHD shell model in the 
$(Pr^{-1}_{\rm M}, Re_{\rm M})$ plane. The dynamo boundary, which
separates dynamo and no-dynamo regions, appears to have a fractal
character. We obtain hysteretic behavior of 
the order parameter across this boundary and suggestions of
nucleation-type phenomena.
\end{abstract}
\maketitle
\section{Introduction\label{intro}}

The elucidation of dynamo action is a problem of central importance in
nonlinear dynamics because it has implications for a variety of physical
systems.  Dynamo instabilities, which amplify weak magnetic fields in a
turbulent conducting fluid, are believed to be the principal mechanism for
the generation of magnetic fields in celestial bodies and in the
interstellar medium~\cite{arnab,rudiger,goedbloed,biskamp,mkvrev,njpspecial,schekochihin}, and in
liquid-metal systems~\cite{roberts,fauve,riga,karl,lathrop,pinton} studied in
laboratories.  In these situations the kinematic viscosity $\nu$ and the
magnetic diffusivity $\eta$ can differ by several orders of magnitude, so
the magnetic Prandtl number $Pr_{\rm M}\equiv\nu/\eta$ can either be very
small or very large; e.g., $Pr_{\rm M}\simeq 10^{-2}$ at the base of the
Sun's convection zone, $Pr_{\rm M}\simeq 10^{-5}$ in the liquid-sodium
system, and $Pr_{\rm M}\simeq 10^{14}$ in the interstellar medium. This
Prandtl number is related to the Reynolds number $Re=UL/\nu$ and the
magnetic Reynolds number $Re_{\rm M}=UL/\eta$ that characterize the
conducting fluid; here $L$ and $U$ are typical length and velocity scales in
the flow; clearly $Pr_{\rm M}=Re_{\rm M}/Re$. 

Two dissipative scales play an important role here; they are the Kolmogorov
scale $\ell_d$ [$\sim \nu^{3/4}$ at the level of Kolmogorov 1941
(K41) phenomenology~\cite{K41}] and the magnetic-resistive scale $\ell_d^M$
[$\sim \eta^{3/4}$ in K41].  For large Prandtl numbers, i.e., $Pr_{\rm M}\gg
1$, $\ell_d^M \ll \ell_d$ so the magnetic field grows predominantly in the
dissipation range of the fluid till it is strong enough to affect the
dynamics of the fluid through the Lorentz force.  This behavior is a
characteristic of a small-scale turbulent dynamo, in which dynamo action is
driven by a smooth, dissipative-scale velocity field.  In the initial stage
of growth, called the kinematic stage of the dynamo, the magnetic
field is not large enough to act back on the velocity field.  Dynamo action
can be obtained for values of $Re_{\rm M}$ that are large enough to overcome
Joule dissipation; and the dynamo-threshold value $Re_{\rm Mb}$ decreases as
$Pr_{\rm M}$ increases~\cite{boldyrev,ponty}.  $Pr_{\rm M} \simeq 10^{-5}$ in
liquid-metal flows~\cite{riga,pinton,kenjeres} so they lie in the
small-Prandtl-number region, $Pr_{\rm M} \ll 1$, for which the growth of the
magnetic energy occurs initially in the {\it inertial scales} of fluid
turbulence, because $\ell_d \ll \ell_d^M$; here the velocity field is not
smooth and the local strain rate is not uniform: At the K41 level the
turnover velocity of an eddy of size $\ell$ is $v(\ell) \sim \ell^{1/3}$, so the
rate of shearing $(v(\ell)/\ell)\sim \ell^{-2/3}$.  

Direct numerical simulations (DNS) are playing an increasingly important role in
developing an understanding of such dynamo action.  Most DNS studies of MHD
turbulence~\cite{frisch,chou,ponty,minnini} have been restricted, because of
computational constraints, either to low resolutions or to the case $Pr_{\rm M}=1$;
small-scale dynamos, with $Pr_{\rm M} \gg 1$ have also been studied via
DNS~\cite{schekochihin}.  However, given the large range spanned by $Pr_{\rm M}$ in the
physical settings mentioned above, some recent DNS studies of the MHD equations have
started to explore the $Pr_{\rm M}$ dependence of dynamo action; the range of $Pr_{\rm
M}$ covered by such pure DNS studies~\cite{ponty,brandenburgdns} is quite modest
($10^{-2}\lesssim Pr_{\rm M} \lesssim 10$). To explore the dynamo boundary in the
$(Pr^{-1}_{\rm M}, Re_{\rm M})$ plane over a large range of $Pr_{\rm M}$, one recent
study~\cite{ponty} has used a combination of numerical methods, some of which require
small-scale models like large-eddy simulations (LES) or lagrangian-averaged MHD (LAMHD),
and others, like a pseudospectral DNS, in which the only approximations are the finite
number of collocation points and the finite step used in time marching; yet another DNS
study~\cite{iskakov} has introduced hyperviscosity of order $8$ to study the low
$Pr_{\rm M}$ regime; by using this combination of methods these studies has been able to
cover the range $10^{-2}\lesssim Pr_{\rm M} \lesssim 10^{3}$ and to obtain the boundary
between dynamo and no-dynamo regions but with fairly large error bars.

We have carried out extensive, high-resolution, numerical studies that have been
designed to explore in detail the boundary between the dynamo and no-dynamo regimes in
the $(Pr^{-1}_{\rm M}, Re_{\rm M})$ plane in a shell model for three-dimensional
MHD~\cite{basu,frick,brandenburgshell,giuliani}.  This shell model allows us to explore
a much larger range of  $Pr_{\rm M}$ than is possible if we use the MHD equations.
Although our study uses a simple shell model, it has the virtue that it can explore the
boundary between dynamo and no-dynamo regions in great detail without resorting to the
modelling of small spatial scales.  Shell-model studies of dynamo action have also been
attempted in Refs.~\cite{frick,plunian,mkvshell,benzi} but these have concentrated on
aspects of the dynamo problem that are different from those we consider here. 

Our study suggests that it is natural to think of the boundary between dynamo and
no-dynamo regimes in the $(Pr^{-1}_{\rm M}, Re_{\rm M})$ plane as a first-order phase
boundary that is the locus of first-order, nonequilibrium phase transitions from one
nonequilibrium statistical steady state (NESS) to another. The first NESS is a
turbulent, but statistically steady, conducting fluid in which the magnetic energy is
negligibly small compared to the kinetic energy; the second NESS is also a statistically
steady turbulent state but one in which the magnetic energy is comparable to the kinetic
energy.  Indeed, the ratio of the magnetic and fluid energies $E_b/E_u$ turns out to be
a convenient order parameter for this nonequilibrium phase transition since it vanishes
in the no-dynamo phase and assumes a finite, nonzero value in the dynamo state. The
other, intriguing result of our study is that the boundary between these phases is very
intricate and might well have a fractal character; this provides an appealing
explanation for the large error bars in earlier attempts to determine this
boundary~\cite{ponty,plunian}.  The analogy with first-order transitions that we have
outlined above is not superficial. As in any first-order transition we find that our
order parameter shows hysteretic behavior~\cite{mrao} as we scan through the dynamo
boundary by changing the forcing term at a nonzero rate. We also find some evidence of
nucleation-type phenomena: the closer we are to the dynamo boundary, the longer it takes
for a significant magnetic field to nucleate and thus lead to dynamo action. We compare
our results with earlier studies such as Ref.~\cite{pontyetal07}, which have suggested
that dynamo action occurs because of a subcritical bifurcation.

The remaining part of this paper is organised as follows: In Sec. \ref{models}
we describe the shell model for MHD~\cite{basu,frick,brandenburgshell} 
and the numerical method we employ. Sec. \ref{results} is devoted to 
our results and Sec. \ref{conclusion} contains a concluding discussion.
  
\section{Models and Numerical Methods\label{models}}

To study dynamo action it is natural to use the equations of 
magnetohydrodynamics(MHD). In three dimensions the MHD equations are
\begin{eqnarray}
\frac{\partial\vec{u}}{\partial t}+(\vec{u}\cdot\nabla)\vec{u} &=&
\nu \nabla^2\vec{u}-\nabla\bar{p}+\frac{1}{4\pi}(\vec{b}\cdot\nabla)
\vec{b}+\vec{f}, \label{magone}\\
\frac{\partial\vec{b}}{\partial t}&=&\nabla\times(\vec{u}\times\vec{b})
+\eta\nabla^2\vec{b}, \label{magtwo}
\end{eqnarray}
where $\nu$ and $\eta$ are the kinematic viscosity and the magnetic
diffusivity, respectively, the effective pressure
$\bar{p}=p+(b^2/8\pi)$, and $p$ is the pressure. For low-Mach-number
flows, to which we restrict ourselves, we use the incompressibility
condition $\nabla\cdot\vec{u}(\vec{x},t) = 0$; and $\nabla
\cdot\vec{b}(\vec{x},t) = 0$.

As we have mentioned above, a DNS of the MHD equations 
poses a significant computational challenge, even on the most powerful
computers available today, if we want to cover a 
large part of the $(Pr^{-1}_{\rm M}, Re_{\rm M})$ plane and to locate the 
dynamo boundary accurately. Therefore one study~\cite{ponty} 
has used a combination of LES, LAMHD, and DNS to obtain this
boundary. We employ a complementary strategy: we use a simple 
shell model for MHD~\cite{basu,frick,brandenburgshell} that
allows us to carry out very extensive numerical simulations to probe 
the nature of the dynamo boundary without using LES or LAMHD. 

Shell models comprise a set of ordinary differential equations with
nonlinear coupling terms that mimic the advection terms in and respect the
shell-model analogs of the conservation laws of the parent hydrodynamic
equations in the inviscid, unforced limit~\cite{bohr,goy}. For the case of
MHD each shell $n$ is characterized by a complex velocity $u_n$ and a
complex magnetic field $b_n$ in a logarithmically discretized Fourier
space with wave vectors $k_n$; furthermore, there is a direct coupling only
between velocities and magnetic fields in nearest and next-nearest neighbor
shells. The MHD shell model equations~\cite{basu,frick} are 
\begin{eqnarray} 
\frac{du_n}{dt}= &-&\nu k_n^2u_n + i[A_n(u_{n+1}u_{n+2}-b_{n+1}b_{n+2})\nonumber \\
&+&B_n(u_{n-1}u_{n+1}-b_{n-1}b_{n+1})\nonumber \label{shell1}\\
&+&C_n(u_{n-2}u_{n-1}-b_{n-2}b_{n-1})]^\ast+f_n^u, \\ 
\frac{db_n}{dt}= &-&\eta k_n^2b_n + i[D_n(u_{n+1}b_{n+2}-b_{n+1}u_{n+2})\nonumber \\
&+&E_n(u_{n-1}b_{n+1}-b_{n-1}u_{n+1})\nonumber \label{shell2}\\
&+&F_n(u_{n-2}b_{n-1}-b_{n-2}u_{n-1})]^\ast+f_n^b, 
\end{eqnarray} 
where $\ast$ denotes complex conjugation, $1\le n \le N$, with $N$ the
total number of shells, the wave numbers $k_n=k_02^n$, with $k_0 = 2^{-4}$,
and $f_n^u$ and $f_n^b$ the forcing terms in the equations for $u_n$ and
$b_n$, respectively.  In our studies of dynamo action, we set $f_n^b=0$.
The parameters $A_n, \, B_n, \ldots , \, F_n$, are obtained by demanding
that these equations conserve all the shell-model analogs of the invariants
of 3DMHD, in the inviscid, unforced case, and reduce to the well-known
Gledzer-Ohkitani-Yamada (GOY) shell model~\cite{bohr,goy} for fluid
turbulence if $b_n = 0$, $\forall~n$.  In particular, to ensure the
conservation of shell-model analogs of the total energy
$E_T=E_u+E_b\equiv(1/2)\sum_n (|u_n|^2 + |b_n|^2)$, cross helicity
$H_C\equiv(1/2)\sum_n(u_nb_n^*+u_n^*b_n)$, and magnetic helicity
$H_M\equiv\sum_n(-1)^n|b_n|^2/k_n$, in the unforced and inviscid case,
and to obtain the GOY-model limit for the fluid, we choose 
\begin{eqnarray} 
A_n&=&k_n;\;\;B_n=-k_{n-1}/2;\;\;C_n=-k_{n-2}/2;\nonumber \\ 
D_n&=&k_n/6;\;\;E_n=k_{n-1}/3;\;\;F_n=-2k_{n-2}/3.  
\end{eqnarray} 
The only adjustable parameters are the forcing terms and $\nu$ and $\eta$.
The ratio $\nu/\eta$ yields the magnetic Prandtl number $Pr_{\rm M}$.
Grashof numbers yield nondimensionalized forces~\cite{prasad} but, for easy
comparison with earlier studies~\cite{schekochihin,ponty,minnini,brandenburgdns,plunian,benzi}, 
we use the fluid and magnetic
integral-scale fluid and magnetic Reynolds numbers whose shell-model analogs
are, respectively,  $Re=u_{\rm rms}\ell_{\rm I}/\nu$, where $\ell_{\rm I} = 
\sum (u_n^2/k_n^2)/\sum (u_n^2/k_n)$ and $u_{\rm rms} =
\sqrt{\sum (u_n^2/k_n)/\ell_{\rm 0}}$, $\ell_{\rm 0} = 2\pi/k_1$
and $Re_{\rm M} = Pr_{\rm M}Re$. 

We use the following boundary conditions: 
\begin{eqnarray} 
A_{N-1}&=&A_N=B_1=B_N=C_1=C_2=0\;;\nonumber \\ 
D_{N-1}&=&D_N=E_1=E_N=F_1=F_2=0.  
\end{eqnarray} 
We set $N=30$ and use a fifth-order,
Adams-Bashforth scheme for solving the shell-model equations,
i.e., for an equation of the type
\begin{equation}
\frac{dq}{dt} = -\alpha q +f(t),
\end{equation}
we use  
\begin{eqnarray}
q(t+\delta t)&=&e^{-2\alpha\delta t}q(t-\delta t)+\frac{1-e^{-2\alpha\delta t}}{
24\alpha}\nonumber \\
&\times&[55f(t)-59f(t-\delta t)\nonumber \\
&+&37f(t-2\delta t)-9f(t-3\delta t)],
\end{eqnarray}
where $\delta t$ is the time step.  We have found that this numerical scheme
works well for the integration of Eqs.(\ref{shell1}) and (\ref{shell2}) so
long as $N \lesssim 35$ and $Re \lesssim 10^9$.  In all our calculations we
use $\delta t = 10^{-4}$.  Characteristic time scales include the time scale
for diffusion $\tau_\eta=\ell_{\rm 0}^2/\eta$ and the large-eddy-turnover time
$\tau_{L}=\ell_{\rm 0}/u_{\rm rms}$, where $\ell_{\rm 0}\equiv 2\pi/k_1$ is the box-size 
length scale and $u_{\rm rms}$ is the root-mean-square velocity.

The initial conditions we use are as follows: We first obtain a statistically steady
state for the GOY-shell-model equations, which are obtained from Eq.(\ref{shell1}) by
setting all $b_n=0$; the forcing terms are chosen to be $f_n^u=f_0(1+i)\delta_{n,1}$,
with $f_0 =  5.0\times 10^{-3}$ in all our runs, except ones in which we study
hysteretic behavior, and $f_n^b=0$. We choose the GOY-model shell velocities at time
$t=0$ to be $u_n = k_n^{-1/3} \exp(i \varphi_n)$, with $\varphi_n$ a random phase
distributed uniformly on the interval $[0,2\pi)$. To make sure we have a statistically
steady state we evolve the shell velocities $u_n$ till $t=5\times10^5$. This yields the
shell-model energy spectrum $E_u(k_n)\equiv | u_n |^2/k_n$ that has the K41 form $\sim
k_n^{-5/3}$ if we ignore intermittency corrections. We now introduce a small seed
magnetic which is such that $E_b\simeq10^{-28}$ and then follow the temporal evolution
of $u_n$ and $b_n$ that is given by Eqs.(\ref{shell1}) and (\ref{shell2}).

\section{Results\label{results}}
\begin{figure}[h!]
\begin{center}
\iffigs
\includegraphics[width=\columnwidth]{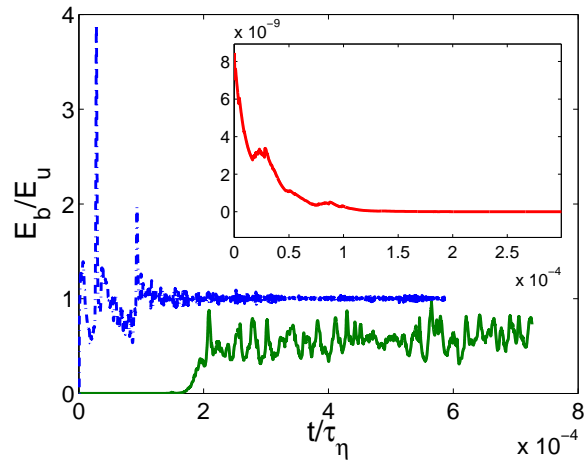}\\
\else\drawing
  50 30 {  } \fi
\end{center}
\vspace{-0.5cm}
\caption[]{\small (Color online) Representative plots of the dynamo order 
parameter $E_b/E_u$ versus time $t/\tau_\eta$, with $\tau_\eta$ the 
magnetic-diffusion
time, in the dynamo region (blue, dashed curve), near the dynamo 
boundary (green, full line), and in the no-dynamo regime (red, full line
in the inset).}
\label{fig:tseries}
\end{figure} 
\begin{figure*}[htb!]
\begin{center}
\iffigs
\includegraphics[width=\textwidth]{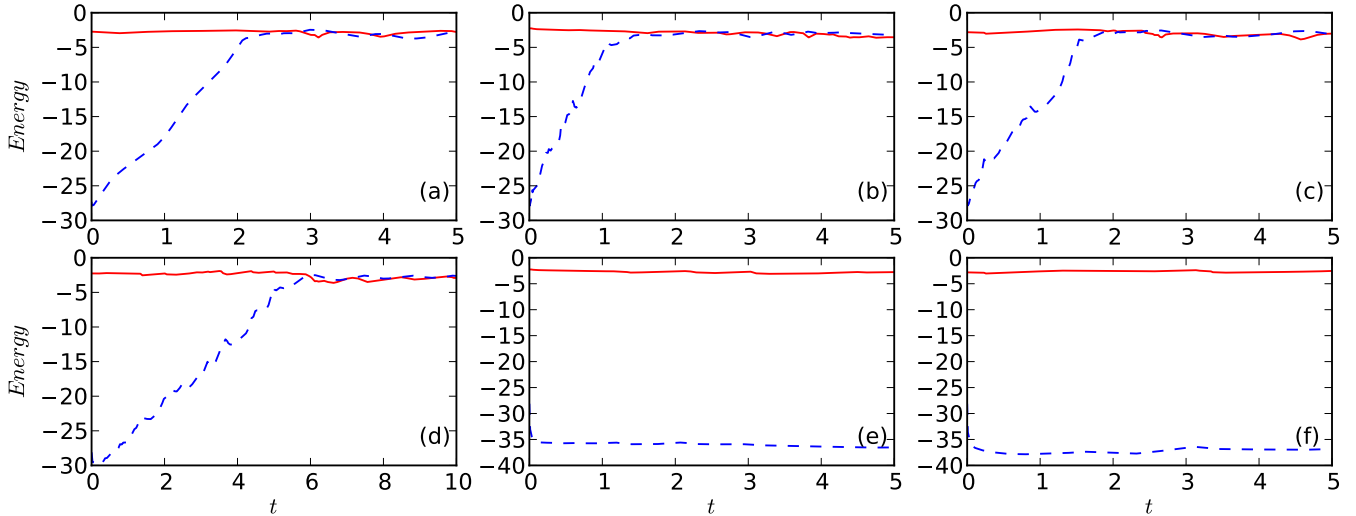}
\else\drawing
  50 30 {  } \fi
\end{center}
\vspace{-0.5cm}
\caption[]{\small (Color online) Semi-log (base 10) plots of
the kinetic (red full curve) and magnetic (blue dashed curve)
energies versus time for (a) $Pr_{\rm M} =10^{2}$, (b) $Pr_{\rm M}=1$, 
(c) $Pr_{\rm M}=10^{-1}$, (d) $Pr_{\rm M}=10^{-2}$, (e) $Pr_{\rm M}=10^{-3}$, 
(f) $Pr_{\rm M}=10^{-6}$. The values of $\nu$ are (a) $10^{-5}$, (b)
$10^{-5}$, (c) $10^{-6}$, (d) $10^{-7}$, (e) $10^{-4}$, and (f) $10^{-7}$; 
and the magnetic-diffusion time $\tau_\eta= \ell_{\rm 0}^2/\eta\simeq
2.52\times10^{10}, ~2.52\times10^{8}, ~2.52\times10^{8}, ~2.52\times10^{4}$ and
$2.52\times10^{4}$, respectively. Dynamo action occurs in 
(a)-(d) but not in (e) and (f).}
\vspace{0.5cm}
\label{fig:energy}
\end{figure*} 
\begin{figure*}[htb!]
\begin{center}
\iffigs
\includegraphics[width=\textwidth]{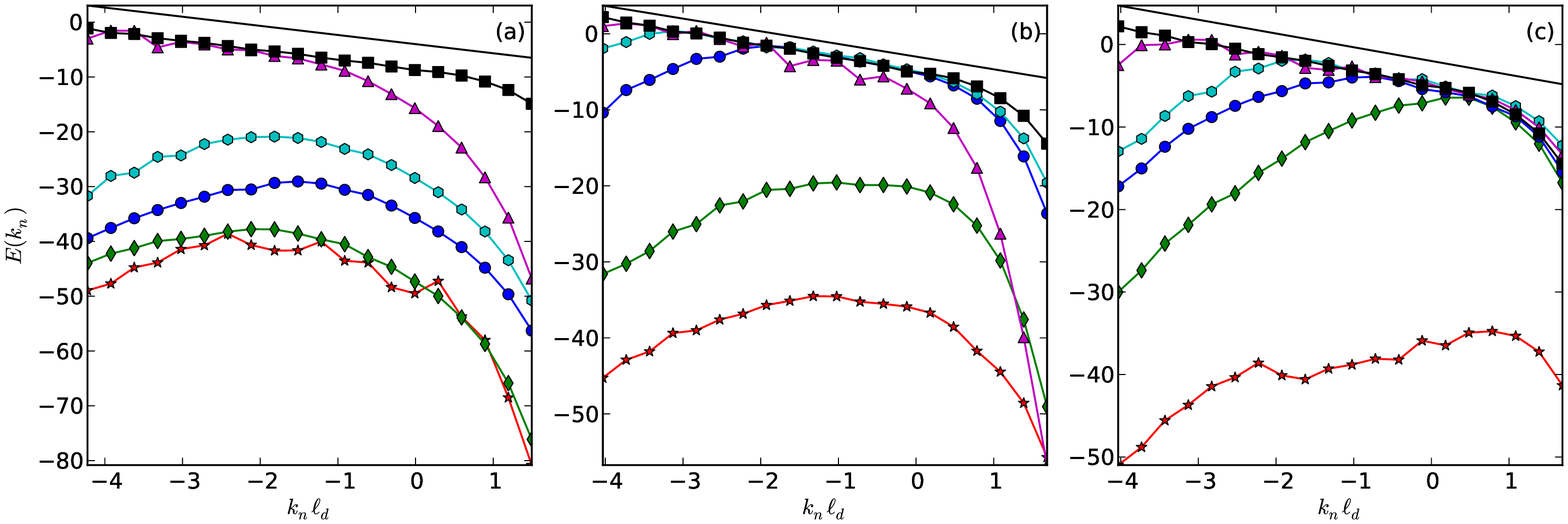}\\
\vspace{-0.35cm}
\includegraphics[width=\textwidth]{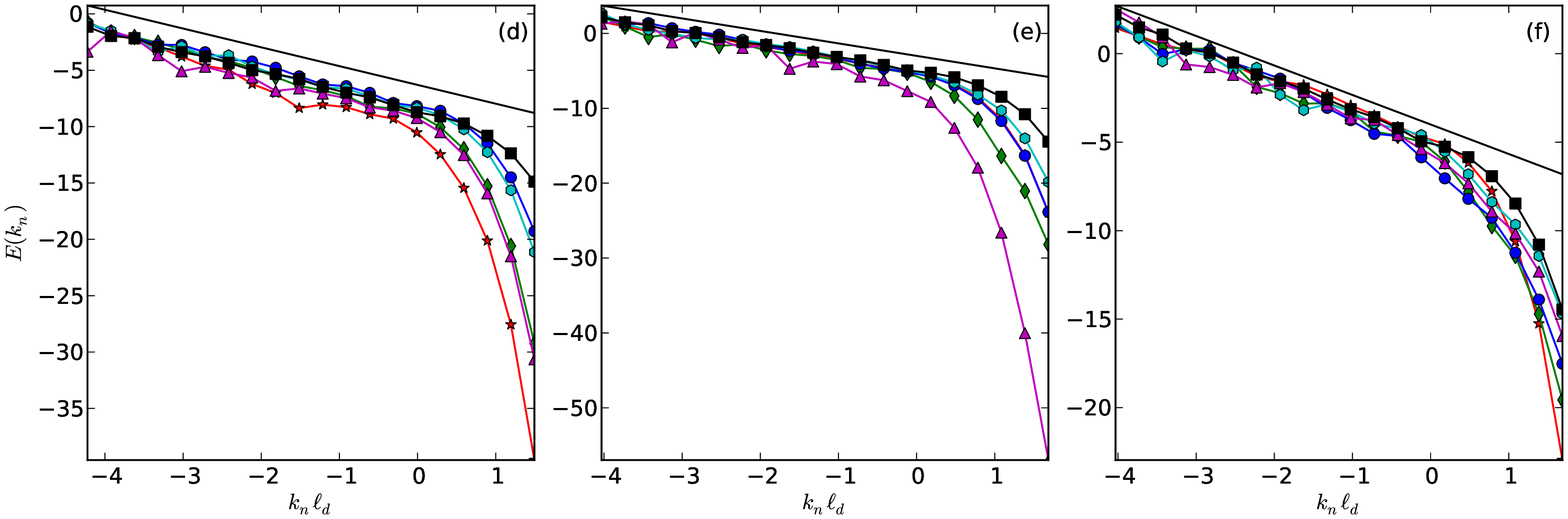}
\else\drawing
  50 30 {  } \fi
\end{center}
\vspace{-0.5cm}
\caption[]{\small (Color online) Log-log (base 10) plots showing
the time evolution of the magnetic-energy spectrum $E_b(k_n)$ 
for representative parameter values at which dynamo action
occurs: (a) $Pr_{\rm M} =10^{-2}, \, \nu = 10^{-7}$; (b) 
$Pr_{\rm M}=1, \, \nu = 10^{-5}$, (c) $Pr_{\rm M}=10^{2} , \, \nu
= 10^{-5}$; analogous plots for kinetic-energy spectra 
are shown in (d), (e), and (f), respectively. The curves with red stars, 
green diamonds, blue hexagons, cyan circles, and magenta triangles, are 
obtained, respectively, at $t = 1, ~5, ~10, ~15,$ and $100$; 
the dissipation scale $\ell_d \simeq 7.422\times10^{-4}$ in (b), (c), (e), and (f); 
$\ell_d \simeq 4.779\times10^{-4}$ in (a) and (d). Curves with black squares
indicate velocity spectra before the seed magnetic field is
introduced; the full black line shows a $k^{-5/3}$ spectrum for
comparison.}
\label{fig:spectra}
\end{figure*} 
\begin{figure}[h]
\begin{center}
\iffigs
\includegraphics[width=\columnwidth]{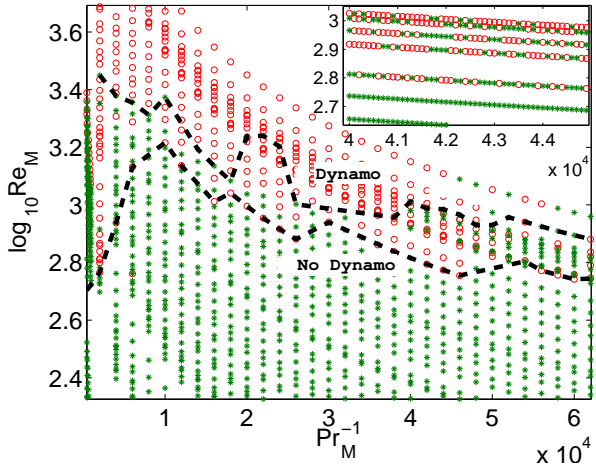}
\else\drawing
  50 30 {  } \fi
\end{center}
\vspace{-0.5cm}
\caption[]{\small (Color online) The dynamo stability diagram in the $(Pr^{-1}_{\rm M}, Re_{\rm M})$
plane: red circles indicate dynamo action; green stars are used if no dynamo occurs. The
boundary between the two regions shows an intricate, interleaved pattern of fine, dynamo
and no-dynamo regimes (see inset for a detailed view). We have drawn two black, dashed
lines; the region above the upper one of these lines is predominantly in the dynamo
regime; the area below the lower one of these lines is principally in the no-dynamo
regime. 
}
\label{fig:stability}
\end{figure} 
\begin{figure*}[htb]
\begin{center}
\iffigs
\includegraphics[width=\columnwidth]{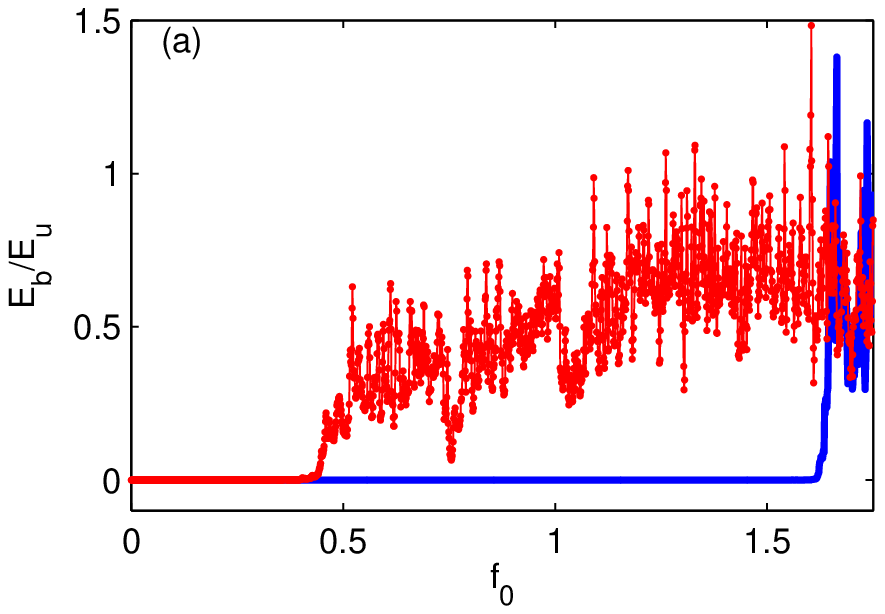}
\includegraphics[width=\columnwidth]{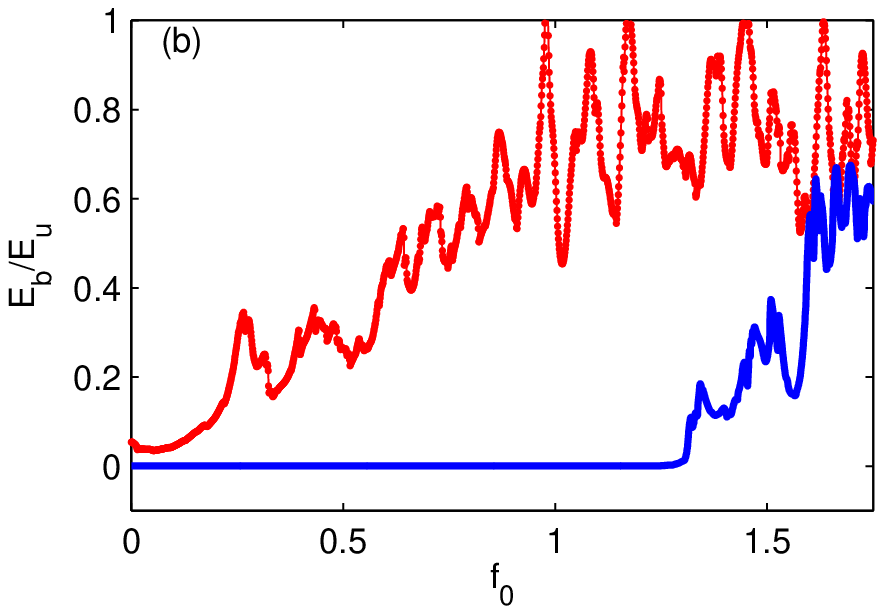}
\else\drawing
  50 30 {  } \fi
\end{center}
\vspace{-0.5cm}
\caption[]{\small (Color online) Plots of the dynamo order parameter $E_b/E_u$ versus the forcing
amplitude $f_0$ illustrating hysteretic behavior as $f_0$ is cycled 
across the dynamo boundary; here $Pr_{\rm M} = 10^{-4}$ and $\nu = 10^{-5}$. As
$f_0$ increases, $E_b/E_u$ follows the blue, full line; if we now decrease $f_0$, then
$E_b/E_u$ follows the red dotted line, and not the blue one, i.e., we have a hysteresis
loop.  We increase $f_0$ in steps of $1.0\times 10^{-3}$ from
an initial value of
$1.0\times 10^{-3}$; we keep $f_0$ constant for a time duration $10$ in (a) and $1$ in
(b); the red, dotted-line segments of the hysteresis loops are obtained by decreasing
$f_0$ at the same rates as for the blue, full-line segments.
}
\label{fig:hysterisis}
\end{figure*} 
\begin{figure*}[htb]
\begin{center}
\iffigs
\includegraphics[width=\columnwidth]{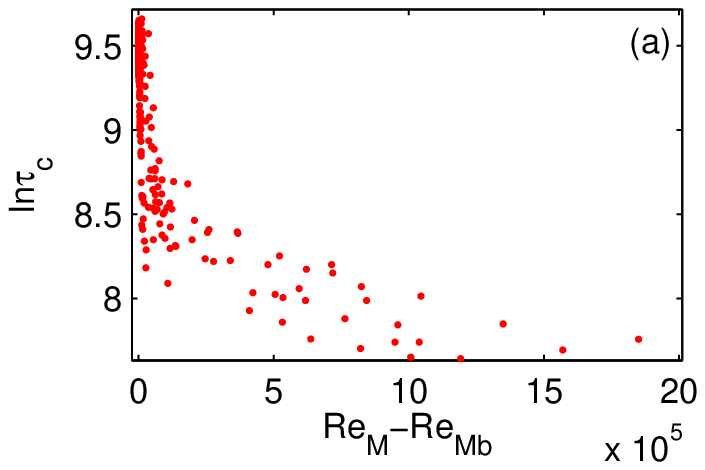}
\includegraphics[width=\columnwidth]{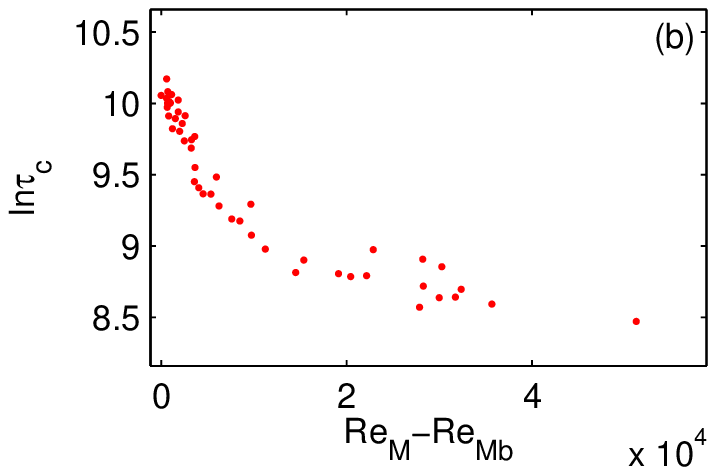}
\else\drawing
  50 30 {  } \fi
\end{center}
\vspace{-0.5cm}
\caption[]{\small (Color online) Representative plots of $\ln \tau_c$
versus $Re_{\rm M}-Re_{\rm Mb}$ for parameter
values at which dynamo action occurs; here $\tau_c$ is
in units of the time step $\delta t$ and $Re_{\rm Mb}$ is
the estimated position of the dynamo boundary:
(a) $Pr_{\rm M}= 1$ and (b) $Pr_{\rm M}=5\times10^{-4}$.
Note that the time $\tau_c$ required for dynamo action increases rapidly
as we approach the dynamo boundary (the plot here is motivated by 
Eq. (27) in Ref.~\cite{oxtoby}).}
\label{fig:time}
\end{figure*}
Given the numerical scheme that we have described in the previous Section, we
obtain the time series for $u_n$ and $b_n$ from the MHD-shell-model
equations. An analysis of these time series shows two types of nonequilibrium
statistical steady states (NESS). We refer to the first as the no-dynamo
state and to the second as the dynamo state. These states have been found in
several earlier studies such as Refs.~\cite{schekochihin,ponty,frisch,chou,
minnini,brandenburgdns,plunian}.
Our main goal is to explore in detail the phase boundary between these two
states. This can be done most easily by the introduction of a dynamo order
parameter a natural candidate for which is the ratio $E_b/E_u$, where the
fluid and magnetic energies are, respectively, $E_u=\frac{1}{2}\sum_n
|u_n|^2$ and $E_b=\frac{1}{2}\sum_n |b_n|^2$.  Representative plots of this
order parameter are given as functions of time $t$ in Fig.~\ref{fig:tseries}:
The blue, dashed curve shows the evolution of $E_b/E_u$ in the dynamo regime;
note that here the dynamo order parameter rises rapidly, fluctuates
significantly for $t/\tau_\eta \lesssim 2 \times 10^{-4}$, and finally
reaches a statistical steady state with equipartition, i.e., $E_b/E_u \simeq
1$.  The red, full curve in the inset of Fig.~\ref{fig:tseries} shows how
$E_b/E_u$ vanishes rapidly in the no-dynamo state. The behavior of the dynamo
order parameter is more complicated than these two simple possibilities in
the vicinity of the phase boundary between dynamo and no-dynamo states as
shown by the green, full curve in Fig.~\ref{fig:tseries}; $E_b/E_u$ rises
much more slowly from zero than in the dynamo regime and then it fluctuates
significantly for a long time; the difficulty of pinpointing the
dynamo boundary is a consequence of these fluctuations. 

The time series for the dynamo order parameter are obtained from those for $E_u$ and
$E_b$; representative plots for these are shown, via red and blue-dashed curves, in
Figs.~\ref{fig:energy}(a), (b), (c), (d), (e), and (f) for a very large range of
magnetic Prandtl numbers, namely, $Pr_{\rm M} = 10^2, ~1, ~10^{-1}, ~10^{-2},
~10^{-3},$ and $10^{-6}$, respectively.  The values of $\nu$ are (a) $10^{-5}$, (b)
$10^{-5}$, (c) $10^{-6}$, (d) $10^{-7}$, (e) $10^{-4}$, and (f) $10^{-7}$; and the
corresponding values of the diffusion time scale $\tau_\eta = \ell_{\rm 0}^2/\eta\simeq
2.52\times10^{10}, ~2.52\times10^{8}, ~2.52\times10^{8}, ~2.52\times10^{4}$ and
$2.52\times10^{4}$, respectively. Clearly dynamo action occurs in
Figs.~\ref{fig:energy}(a)-(d) but not Figs.~\ref{fig:energy}(e) and (f).  By obtaining
many such plots we can identify the dynamo boundary in the $(Pr^{-1}_{\rm M}, Re_{\rm
M})$ plane as we discuss later.

In the dynamo regime the shell-model kinetic and magnetic energy spectra
defined, respectively, by  $E_u(k_n)\equiv |u_n|^2/k_n$ and $E_b(k_n)\equiv
|b_n|^2/k_n$ evolve as shown in Fig.~\ref{fig:spectra}. In particular, in
Figs.~\ref{fig:spectra}(a), (b), and (c) for $Pr_{\rm M} = 10^{-2}, ~1,$
and $10^2$, respectively, we show the evolution of $E_b(k_n)$ with time: the
curves with red stars, green diamonds, blue hexagons, cyan circles, and
magenta triangles, are obtained, respectively, for $t = 1, ~5, ~10, ~15,$
and $100$; the analogs of these plots for $E_u(k_n)$ are given in
Figs.~\ref{fig:spectra}(d), (e), and (f). Note that the initial growth of
$E_b(k_n)$ occurs principally at large values of $k_n$ if $Pr_{\rm M}$ is
large, i.e., we have a small-scale dynamo; this growth of $E_b(k_n)$ moves to
low values of $k_n$ as $Pr_{\rm M}$ decreases; earlier 
studies~\cite{chou,plunian} have observed similar trends but not over 
the large range of $Pr_{\rm M}$ we cover. As $E_b(k_n)$ grows, the velocity 
spectra are also affected but much less than their magnetic counterparts as
can be seen by comparing Figs.~\ref{fig:spectra}(d), (e), and (f) with 
Figs.~\ref{fig:spectra}(a), (b), and (c), respectively. In all these plots
the curves with black squares indicate $E_u(k_n)$ from the initial
steady state for the GOY shell model; and the black lines with no symbols
show the K41 $k_n^{-5/3}$ spectrum for comparison. From this line we see
that the NESS that is obtained, once dynamo action has occurred, is such that
both velocity and magnetic-field energy spectra display a 
substantial inertial range with K41 scaling; these inertial ranges are
not large enough, at least near the dynamo boundary in our runs, for
a reliable estimation of multiscaling corrections to the $-5/3$
exponent. If $Pr_{\rm M} \simeq 1$ then the scaling ranges in velocity and
magnetic-field spectra are comparable; as $Pr_{\rm M}$
decreases (increases), the scaling range for the magnetic spectrum 
decreases (increases) relative to its counterpart in the velocity
spectrum; these trends are clearly visible in the 
representative plots in Fig.~\ref{fig:spectra}.

We return now to the identification of the dynamo boundary.  A close scrutiny of the
plots in Fig.~\ref{fig:energy} shows that the initial growth of $E_b$ is not monotonic.
It is important, therefore, to set a threshold value of the magnetic energy $E_b^c$: For
a given pair of values for $Pr_{\rm M}$ and $Re_{\rm M}$, if $E_b(t) > E_b^c$ for $t >
\tau_c$, where $\tau_c$ is the time at which the threshold value is crossed, we conclude
that dynamo action occurs; if not, then there is no dynamo formation.  By examining the
growth of $E_b(t)$ we can, therefore, map out the dynamo boundary in the $(Pr^{-1}_{\rm
M}, Re_{\rm M})$ plane.  The crossing time $\tau_c$ depends on $Pr_{\rm M}$ and $Re_{\rm
M}$. Note that, if $\tau_c(Pr_{\rm M}, Re_{\rm M}) > t_{\rm max}$, the length of time
for which we integrate Eqs.(\ref{shell1}) and (\ref{shell2}), we would conclude, {\it
incorrectly}, that no dynamo action occurs for this of values of $Pr_{\rm M}$ and
$Re_{\rm M}$.  In other words the dynamo boundary depends on $t_{\rm max}$;  we have
checked this explicitly in several cases.

An important questions arises now: Is there a well-defined dynamo boundary in the
$(Pr^{-1}_{\rm M}, Re_{\rm M})$ plane as $t_{\rm max}\rightarrow \infty$?  Earlier
studies~\cite{schekochihin,ponty,minnini} have began to answer this question. They find that, if
$t_{\rm max}\simeq\tau_\eta$, then a well-defined dynamo boundary is obtained. However,
since they work with the MHD equations the error bars on this boundary are large and the
range of values of $Pr_{\rm M}$ and $Re_{\rm M}$ rather limited.  

The simplicity of our model allows us to carry out a systematic study of the dynamo
boundary.  We find that, at least in our shell model for MHD, we can obtain an
asymptotic dynamo boundary (see Fig.~\ref{fig:stability}) if we choose $E_b^c = 0.9
E_u$, i.e., we conclude that dynamo action has occurred if $E_b(t)$ exceeds $0.9 E_u$;
furthermore, if $E_b(t)$ falls below $10^{-35}$ we say that dynamo action will never be
achieved. We continue the temporal evolution of Eqs.(\ref{shell1}) and (\ref{shell2})
till one of these criteria is satisfied.  For all values of $Pr_{\rm M}$ and $Re_{\rm
M}$ that we have used we find that this $t_{\rm max}$, the run time required to decide
whether or not dynamo action occurs, is several orders of magnitude lower than
$\tau_\eta$.  We have also checked for several representative pairs of values for
$Pr_{\rm M}$ and $Re_{\rm M}$ that runs of length $t_{\rm max}\simeq\tau_\eta$ do not change
our conclusions about such dynamo action.

The dynamo boundary that we obtain is shown in the stability 
diagram of Fig.~\ref{fig:stability}. Red circles indicate parameter
values at which we obtain dynamo action whereas green stars are used
for values at which no dynamo occurs. The most important result that 
follows from this stability diagram is that the boundary between dynamo
and no-dynamo regimes is very complicated. It seems to be 
of fractal-type, with an intricate pattern of fine, dynamo regions
interleaved with no-dynamo regimes. This is especially apparent in the inset
of Fig.~\ref{fig:stability}, which shows a detailed view of the stability 
diagram in the vicinity of the dynamo boundary. Earlier studies 
seem to have missed this fractal-type of boundary because they have
not been able to examine the transition in as much detail as we
have for our shell model. However, fractal-type boundaries between different
dynamical regimes have been suggested in other extended dynamical systems;
recent examples include the transition to turbulence in pipe flow~\cite{bruno}
and different forms of spiral-wave dynamics in mathematical models
for cardiac tissue~\cite{shajahan}. In Fig.~\ref{fig:stability} we have drawn
two black, dashed lines; the region above the upper one of these lines is 
predominantly in the dynamo regime; the area below the lower one of these
lines is predominantly in the no-dynamo regime. These two lines give an
approximate indication of the error bars we might expect in the determination
of the dynamo boundary in a study that cannot scan through points in
the $(Pr_{\rm M}^{-1}, Re_{\rm M})$ plane as finely as we have.

From Fig.~\ref{fig:tseries} we see that the order parameter $E_b/E_u$ jumps from a very
small value in the no dynamo region to a value $\simeq 1$ in the dynamo state. It is
natural, therefore, to think of the dynamo boundary as a nonequilibrium, first-order
boundary. In an equilibrium, first-order transition the order parameter shows hysteretic
behavior if we scan through a first-order boundary by, say, changing, at a finite rate,
the field that is conjugate to the order parameter~\cite{mrao}. It is natural to ask if
we see such hysteretic behavior at the dynamo boundary. Indeed, we do, as we show in
Fig.~\ref{fig:hysterisis} where we cross the dynamo boundary by changing the amplitude
$f_0$ of the forcing term in Eq.(\ref{shell1}).  Figure~\ref{fig:hysterisis} shows
representative plots of the dynamo order parameter $E_b/E_u$ versus $f_0$; these
illustrate the hysteretic behavior that occurs when $f_0$ is cycled at a finite, nonzero
rate across the dynamo boundary; here $Pr_{\rm M} = 10^{-4}$ and $\nu = 10^{-5}$. As
$f_0$ increases, $E_b/E_u$ follows the blue, full line: it increases and then saturates;
fluctuations are superimposed on these mean trends. If we now decrease $f_0$, then
$E_b/E_u$ follows the red dotted line, and not the blue one, i.e., we have a hysteresis
loop. The faster the rate at which we change $f_0$ the wider is the hysteresis loop as
is known from studies of hysteresis in spin systems~\cite{mrao}. Here we increase $f_0$
in steps of $1.0\times 10^{-3}$ from an initial value of $1.0\times 10^{-3}$; we keep
$f_0$ constant for a time period $10$ in Fig.~\ref{fig:hysterisis}(a) and $1$ in
Fig.~\ref{fig:hysterisis}(b); the red, dotted-line segments of the hysteresis loops are
obtained by decreasing $f_0$ at the same rates as for the blue, full-line segments; the
loop in the former case is narrower than in the latter.

Given the analogy with first-order transitions that we have outlined above,
it is natural to ask if nucleation-type phenomena~\cite{oxtoby} are also
associated with dynamo formation. It would be interesting to check this in a 
DNS of the MHD equations. At the level of our shell model, the best we can do 
is to try to see if, for a given $Pr_{\rm M}$, when we obtain a dynamo, 
the time required for dynamo action $\tau_c$ diverges as we approach the 
dynamo boundary. Our data are consistent with an increase of $\tau_c$ 
as we approach this boundary from the dynamo side as shown by the 
representative plots in Fig.~\ref{fig:time}. However, it is hard to fit a precise
form to the behavior of $\tau_c$ near the dynamo boundary 
partly because of the complicated nature of this boundary
which makes it difficult to estimate the position $Re_{\rm Mb}$
reliably (the plot in Fig.~\ref{fig:time} is motivated by the form of 
Eq. (27) in Ref.~\cite{oxtoby}).

\section{Conclusions\label{conclusion}}

We have presented a detailed study of dynamo action in a shell model of
turbulence~\cite{basu,frick,brandenburgshell}. Our study has been designed to explore
the nature of the boundary between dynamo and no-dynamo regimes in the $(Pr^{-1}_{\rm
M}, Re_{\rm M})$ plane over a much wider range of $Pr_{\rm M}$ than has been attempted
in earlier numerical studies. The dynamo boundary emerges as a first-order
nonequilibrium phase boundary between one turbulent, nonequilibrium statistical steady
state (NESS) and another~\cite{footnote}.
This point of view is implicit in earlier work, e.g., in
studies of the Kazantsev dynamo~\cite{kazantsev} or in studies that view dynamo
generation as a subcritical bifurcation~\cite{busse77,pontyetal07,kuangetal08}.  One of these
studies~\cite{pontyetal07} has remarked that when dynamo action `` ... is obtained in a
fully turbulent system, where fluctuations are of the same order of magnitude as the
mean flow ... the traditional concept of amplitude equation may be ill-defined and
one may have to generalize the notion of ``subcritical transition" for turbulent flows
...". We believe that the natural generalization is the nonequilibrium, first-order
transition we suggest above. We have explored the explicit consequences of such a view
in far greater detail than has been attempted hitherto. In particular, the ratio
$E_b/E_u$ is a convenient order parameter for this nonequilibrium phase transition; it
shows hysteresis across the dynamo boundary like order parameters at any first-order
transition; and nucleation-type phenomena also seem to be associated with dynamo
formation.  Last, and perhaps most interesting, we find that the dynamo boundary seems
to have a fractal character; this provides a natural explanation for the large error
bars in earlier attempts to determine this boundary~\cite{schekochihin,ponty,plunian}.
Furthermore, this fractal-type boundary might well be the root cause of magnetic-field
reversals discussed, e.g., in Refs.~\cite{benzi,pintonreversal}.  

It is important to check, of course, that our shell-model results carry over to the MHD
equations.  This requires large-scale DNS that might well be beyond present-day
computing capabilities if we want to explore issues like the possible fractal nature of
the dynamo boundary. However, analogs of the hysteretic behavior we mention above have
been obtained in DNS studies of the MHD
equations~\cite{busse77,pontyetal07,kuangetal08}; hysteresis has also been seen in a
numerical simulation that includes turbulent convection~\cite{sim+bus09}. In some of
these studies hysteretic behavior is obtained by changing the viscosity of the magnetic
Prandtl number. We have obtained hysteresis by changing the forcing; this change of
forcing might be easier to effect in experiments than a change of the viscosity or
magnetic diffusivity.
 
To the best of our knowledge, earlier studies have not noted the increase in the 
dynamo-formation time $\tau_c$ as the dynamo boundary is approached from the
dynamo side. We have suggested that this is akin to the increase in the time
required to form a critical nucleus as we approach a first-order boundary~\cite{oxtoby}.
It would be interesting to see if such an increase of $\tau_c$ can be obtained
in DNS studies of dynamo formation with the MHD equations. It is worth noting here 
that some DNS studies~\cite{ponty} have suggested that simulation times comparable 
to the diffusion time scale $\tau_\eta$ are required to confirm dynamo formation;
by contrast our shell-model study yields dynamo action on a much shorter time $\tau_c$,
which increases as we approach the dynamo boundary. Perhaps the large simulation
times required for dynamo action in full MHD simulation might have arisen because 
these simulations have been carried out in the vicinity of the dynamo boundary.

To settle completely whether the dynamo boundary is of fractal-type, very long
simulations might be required to make sure that the apparent fractal nature is not an
artifact of long-lived metastable states. To make sure that our calculations do not
suffer from such an artifact, we have carried out very long runs for representative
points in the region of the dynamo boundary in Fig.~\ref{fig:stability}; we have found
that these long runs do not change our results. Furthermore, it is useful to check
whether, instead of one dynamo boundary, there is a sequence of transitions, with more
and more complicated temporal behaviors for the order parameter, as has been seen in the
turbulence-induced melting of a nonequilibrium vortex crystal~\cite{crystalmelt}. We
have not found any conclusive evidence for this but, in the vicinity of the dynamo
boundary, the order parameter can oscillate for fairly long times (see, e.g., the green
full curve in Fig.~\ref{fig:tseries}). To decide conclusively whether these oscillations
characterize a new nonequilibrium oscillating state, different from the simple dynamo 
and no-dynamo NESSs we have mentioned, requires extensive numerical studies that lie 
beyond the scope of this paper.

In equilibrium statistical mechanics different ensembles are equivalent; in particular,
we may determine a first-order phase boundary by using either the canonical or the
grand-canonical ensemble. However, such an equivalence of ensembles does not apply to
transitions between different nonequilibrium statistical steady states (NESSs); examples
may be found in driven diffusive systems~\cite{acharyya} or in the turbulence-induced
melting of a nonequilibrium vortex crystal~\cite{crystalmelt}.  Given that the dynamo
boundary separates two turbulent NESSs, we might expect that this boundary might depend
on precisely how the system is forced.  Evidence for this exists already: For example,
the dynamo boundary depends on whether a stochastic external force is
used~\cite{schekochihin} or whether a Taylor-Green force is used~\cite{ponty};
furthermore, this boundary is different if the fluid is helical~\cite{brandenburgdns},
as in most astrophysical dynamos. 

We hope our study of dynamo formation in a shell model for MHD will stimulate both DNS
and experimental studies designed to explore the first-order nature of the dynamo
transition.

\begin{acknowledgments}
We thank R. Karan, S.S. Ray, and S. Ramaswamy for discussions, 
SERC(IISc) for computational resources and DST, UGC and CSIR India 
for support. One of us is a member of the International Collaboration 
for Turbulence Research (ICTR).
\end{acknowledgments}


\begin{thebibliography}{50}
\bibitem{footnote} We use the expression first-order transition as in 
equilibrium statistical mechanics since the order parameter (the first 
derivative of a free energy) jumps at the transition. Since there is no 
free energy for the NESSs we consider, one can, just as well,
refer to this as a discontinuos transition.
\end{thebibliography}
\end{document}